\documentclass[aps,preprint,amsmath,amssymb,nofootinbib]{revtex4-1}
\usepackage{graphicx} 
\usepackage{amsmath}
\usepackage{amsfonts,amsbsy}
\usepackage{amssymb}

\def\gsim{ \,\, \vcenter{\hbox{$\buildrel{\displaystyle >}\over\sim$}}
 \,\,}

\def\be{\begin{equation}}
\def\ee{\end{equation}}
\def\bea{\begin{eqnarray}}
\def\eea{\end{eqnarray}}
\def\tr{{\rm tr}\,}

\begin{document}

\title{\bf Magnetic screening in high-energy heavy-ion collisions}

\preprint{RBRC 1020}

\author{Adrian Dumitru$^{a,b,c}$, Hirotsugu Fujii$^d$ and Yasushi Nara$^{e}$}
\affiliation{
$^a$ RIKEN BNL Research Center, Brookhaven National
  Laboratory, Upton, NY 11973, USA\\
$^b$ Department of Natural Sciences, Baruch College, CUNY,
17 Lexington Avenue, New York, NY 10010, USA\\
$^c$ The Graduate School and University Center, The City
  University of New York, 365 Fifth Avenue, New York, NY 10016, USA\\
$^d$ Institute of Physics, University of Tokyo, Komaba, Tokyo 153-8902, Japan\\
$^e$ Akita International University, Yuwa, Akita-city 010-1292, Japan\\
}

\begin{abstract}
We show that classical chromomagnetic fields produced coherently in
the initial stage of a heavy-ion collision exhibit screening. From the
two-point field strength correlator we determine the magnetic mass for
SU(2) to be $m_M\simeq 5$ times the saturation scale. Magnetic
screening leads to an intuitive understanding of the area law scaling
of spatial Wilson loops observed previously. The presence of screening
effects in the initial state provides a basis for defining kinetic
processes in the early stage of heavy-ion collisions, with electric
and magnetic masses of the same order.
\end{abstract}

\maketitle

Heavy ion collisions at high energies involve non-linear dynamics of
strong QCD color fields~\cite{Mueller:1999wm}. The soft field of a
dense system of color charges at rapidities far from the source,
resp.\ at light-cone momentum fractions $x\ll1$, is determined by the
classical Yang-Mills equations with a static current on the light
cone~\cite{MV}. It consists of gluons with a transverse momentum on
the order of the density of valence charges per unit transverse area,
$Q_s^2$~\cite{JalilianMarian:1996xn}. {\em Parametrically}, the
  saturation momentum scale $Q_s$ separates the regime of non-linear
color field interactions from the perturbative (linear) regime.
$Q_s$ is commonly defined from a two-point function of {\em
  electric} Wilson lines, the ``dipole scattering amplitude'' ${\cal
  N}(r)$, evaluated in the field of a single hadron or
nucleus~\cite{Kovchegov:1998bi}; in covariant gauge
\bea
V({\bf x}) &=& {\cal P} \, \exp\left( ig\int dx^- A^+(x^-,{\bf x})\right) ~,\\
{\cal N}(r) &=& \left< 1-\tr V({\bf 0}) V^\dagger({\bf x})\right>
\simeq 1 - \exp\left( -\frac{1}{4} \; r^2\; Q_s^2\right)~.
\eea
We refer to refs.~\cite{p_Pb_predict} for summaries of recent
predictions specifically for the p+Pb collision run at the LHC where a
dense nucleus is probed by a dilute projectile.

The soft field produced in a collision of two dense nuclei is then
obtained from the classical Yang-Mills equations subject to
appropriate matching conditions on the light
cone~\cite{Kovner:1995ja}. Right after impact longitudinal
chromo-electric and magnetic fields $E_z,~B_z\sim 1/g$ dominate; there
is a source for $B_z$ because the projectile and target fields do not
commute~\cite{Fries:2006pv,TL_LM}. They fluctuate according to the
random local color charge densities of the valence sources described
by a quadratic effective action
\be \label{eq:S2}
S_{\rm eff}[\rho^a] = \frac{\rho^a({\bf x}) \rho^a({\bf
    x})}{2\mu^2}~~~,~~~ \langle \rho^a({\bf x})\, \rho^b({\bf
  y})\rangle =\mu^2 \delta^{ab} \delta({\bf x}-{\bf y})~, 
\ee 
with $\mu^2$ proportional to the thickness of a given
nucleus~\cite{MV}. The width of Gaussian color charge fluctuations
also sets the saturation scale: $Q_s^2 \sim g^4 \mu^2$.

Before the collision the individual fields of projectile and target
are 2d pure gauges,
\be \label{eq:alphai} 
\alpha^i_m = \frac{i}{g} \, U_m \, \partial^i
U_m^\dagger~~~~,~~~~ \partial^i \alpha^i_m = g \rho_m~,
\ee
where $m=1,\, 2$ labels projectile and target, respectively, and $U_m$
are SU(N) fields. Eqs.~(\ref{eq:alphai}) can be solved either
analytically in an expansion in the charge density / field
strength~\cite{Kovner:1995ja} or numerically on a two-dimensional
lattice~\cite{Krasnitz:1998ns,CYMlatt}. We restrict here to a single
rapidity slice and so do not consider a longitudinally extended
source~\cite{Fukushima:2007ki}; the parameter $\mu^2$ in~(\ref{eq:S2})
is to be understood as integrated over the thickness of the source in
rapidity.

The field in the forward light cone immediately after the collision,
at proper time $\tau\equiv\sqrt{t^2-z^2}\to0$, is given by
$A^i=\alpha_1^i + \alpha_2^i$~\cite{Kovner:1995ja} in the
continuum\footnote{See ref.~\cite{Krasnitz:1998ns} for the
  corresponding expressions on the lattice.}. This leads
to~\cite{TL_LM}
\be
F_{xy} = ig\epsilon^{ij} \left[\alpha_1^i , \alpha_2^j\right]~~~~,~~~~
{\bf \nabla\cdot B} = ig \left[A^i, B^i \right]~.
\ee
Note the presence of sources / sinks for magnetic field lines.

It has been shown recently~\cite{Dumitru:2013koh} that spatial Wilson
loops in the field $A^i$ of produced soft gluons satisfy area law
scaling for areas $\gsim 1.5/Q_s^2$. To confirm that this is due to
screening of magnetic fields we consider here the two-point
correlator\footnote{We denote $\tr$ the trace over fields in the
  fundamental representation, divided by the dimension of that
  representation (i.e., by the number of colors $N_c=2$).}  of the
longitudinal magnetic field strength $F_{xy}$
\bea
C^{(2)}(r) &=& g^2 \left< \tr G_{xy}({\bf 0})\,  G_{xy}({\bf x})
\right>~, \label{eq:C2}\\
G_{xy}({\bf x}) &\equiv& U_{\bf 0\to x} \, F_{xy}({\bf x})\,  U^{-1}_{\bf 0\to x}~.
\label{eq:C2b}
\eea
In the second line we perform a parallel transport of the gluon field
to the origin, i.e.\ $U_{\bf 0\to x}$ denotes a product of the links
along some path\footnote{We sum over the two paths along the sides of
  a rectangle in the plane $z=0$ with ${\bf 0}$ and ${\bf x}$ on
  diagonally opposed corners. For any particular configuration of
  links $G_{xy}({\bf x})$ obviously is a function of the choice of
  path; the ensemble averaged correlator $C^{(2)}$ is independent of
  the path.} from ${\bf 0\to x}$. The ensemble average $\langle\cdot
\rangle$ in eq.~(\ref{eq:C2}) is performed with a sum of two actions
like in eq.~(\ref{eq:S2}); they describe valence charge fluctuations
of projectile and target, respectively.

We emphasize that the propagator $C^{(2)}(r)$ is different from
\be \label{eq:BB}
\left< \tr F_{xy}({\bf 0})\,  F_{xy}({\bf x}) \right>~.
\ee
The latter is not gauge invariant. In~(\ref{eq:BB}) the external legs
interact only with the light-cone sources (the two-dimensional pure
gauges) while the parallel transporters in~(\ref{eq:C2},\ref{eq:C2b})
introduce interactions with the produced background field. Hence, the
behavior of~(\ref{eq:BB}), for which an explicit expression is given
in ref.~\cite{Fujii:2008km}, differs from that of $C^{(2)}(r)$.

The correlation function $C^{(2)}(r)$ is gauge invariant and shall be
used below to define the magnetic screening scale via
\be \label{eq:C2fit}
C^{(2)}(r) \sim \frac{1}{\sqrt{m_M\; r}} \, \exp\left( - m_M\; r\right)~.
\ee
This form corresponds to a screened propagator in $d=2$ dimensions,
\be
\int d^dp \, \frac{1}{p^2+m^2} \, e^{-i {\bf p \cdot x}}
\sim \frac{1}{(m\, r)^{(d-1)/2}} \, e^{-m\, r}~~~~~~~~~(m\, r\gsim 1).
\ee
We should mention here that in the present setup screening masses
should not be sensitive to ultraviolet
cutoffs~\cite{Bodeker:1995pp}. This is due to the fact that the phase
space density of gluons drops like $\sim 1/k_\perp^4$ at high
momentum~\cite{Kovner:1995ja}, much more rapidly than for a classical
field in thermal equilibrium ($\sim 1/|{\bf k}|$). Furthermore, due to
the fact that the occupation number of the classical field is ${\cal
  O}(1/g^2)$, the screening mass extracted from the
correlator~(\ref{eq:C2}) is $m^2/Q_s^2 = {\cal O}(g^0)$.

One may generalize eq.~(\ref{eq:C2}) to higher point functions such
as
\be
C^{(3)}({\bf x,y}) = g^3 \left< \tr G_{xy}({\bf 0})\,  G_{xy}({\bf x})
\,  G_{xy}({\bf y}) \right>~.    \label{eq:C3}
\ee
However, the operator on the rhs of this equation is odd under charge
conjugation ${\cal C}$ of the gauge field while the
action~(\ref{eq:S2}) is ${\cal C}$-even. Hence, for that action
$C^{(3)}=0$. Non-zero $C^{(3)}$ could be obtained for three or more
colors by adding the ``odderon'' operator $\sim d^{abc} \rho^a \rho^b
\rho^c$ to~(\ref{eq:S2})~\cite{rho_n_action}. Here we only consider
SU(2) gauge fields.

The magnetic two-point correlator~(\ref{eq:C2}) is computed on a
two-dimensional periodic lattice with $N_s$ sites per dimension, for a
given value of $\mu_L\equiv g^2\mu a$, where $a$ is the lattice
spacing. The continuum limit is approached as $\mu_L\to0$ which
widens the correlation function $C^{(2)}$ over an increasing number
of lattice sites. At the same time, to ensure that finite size effects
are small we choose $\mu_L N_s\gsim 100$.

\begin{figure}[htb]
\begin{center}
\includegraphics[width=0.66\textwidth,angle=270]{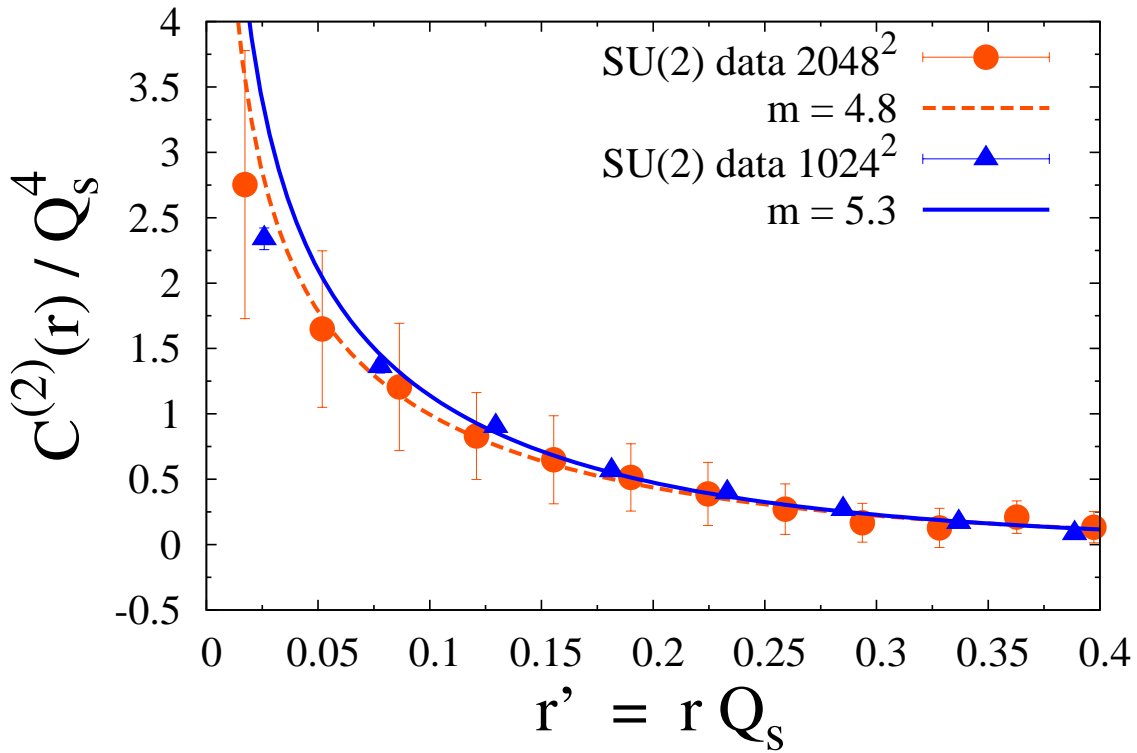}
\end{center}
\vspace*{-0.4cm}
\caption[a]{Two point magnetic field strength correlator $C^{(2)}(r)$.
  We define $Q_s^2 = (C_F/2\pi)\,g^4 \mu^2$. Symbols show numerical
  results for $SU(2)$ Yang-Mills on $2048^2$ ($1024^2$) lattices at
  $\mu_L \equiv g^2\mu a=0.1$ ($\mu_L=0.15$). The lines represent fits
  (excl.\ the first data point) of the form $C^{(2)}(r)\sim \exp(-m\;
  r)/\sqrt{m\; r}$. The extracted magnetic mass is $m_M/Q_s\approx 5\pm
  0.5$.}
\label{fig:C2}
\end{figure}
Fig.~\ref{fig:C2} shows the correlator of magnetic field strengths as
a function of distance. The lines represent fits of the
form~(\ref{eq:C2fit}). The first data point from either set was
excluded from the fit as one might expect lattice discretization
effects to be large; also, eq.~(\ref{eq:C2fit}) receives corrections
at small $m r$. The fit estimates an asymptotic standard error for
the mass parameter of about 10\%.

The numerical data matches the 2d screened propagator rather well. We
extract a surprisingly large value for the magnetic mass, $m_M\simeq
(5 \pm 0.5) \; Q_s$, which shows that the theory linearizes only on
momentum scales quite a bit beyond $Q_s$. In simple terms, screening
arises because of the presence of effective magnetic charges mentioned
above; therefore, magnetic field lines do not escape to infinity.

Using the fit to $C^{(2)}(r)$ (incl.\ proper normalization) in the
relation~\cite{Di Giacomo:2000va}
\be
\sigma_M = \frac{1}{2} \int d^2r \, C^{(2)}(r)   \label{eq:sigma_C2}
\ee
reproduces the spatial string tension $\sigma_M \simeq 0.12 \; Q_s^2$
obtained in ref.~\cite{Dumitru:2013koh} to about
15\%. Eq.~(\ref{eq:sigma_C2}) arises in a cluster expansion of the
Wilson loop~\cite{Di Giacomo:2000va}. The crucial point is
that~(\ref{eq:sigma_C2}) is {\em independent} of the area over which
one integrates as long as magnetic fields are screened over
significantly shorter scales. A generalization to higher-order
cumulants of field strength correlators~\cite{Di Giacomo:2000va} would
then explain the area law scaling of spatial Wilson loops observed
previously~\cite{Dumitru:2013koh}. Furthermore, it would
be very interesting to determine the time evolution of $C^{(2)}$ from
$\tau=0$ to $\tau\sim1/Q_s$. Because of the rather large value of the
mass it is a difficult task to ensure that on the lattice the fields
propagate with (nearly) the continuum-limit frequency.

In summary, we have shown that the chromomagnetic field produced
coherently in a high-energy collision of dense color charges exhibits
screening. Magnetic field lines do not escape to infinity but are
captured by effective sources of non-Abelian magnetic flux. We obtain
a rather large magnetic mass of about 5 times the saturation scale
$Q_s$ (for $N_c=2$ colors). Thus, in a heavy-ion collision ``naive''
(unscreened) perturbation theory applies only well beyond $Q_s$. The
rather short screening length provides an intuitive interpretation for
the onset of area law scaling of the spatial Wilson loop already for
radii $R \sim 0.8/ Q_s$~\cite{Dumitru:2013koh}. Finally, the presence
of screening effects in the initial state should be relevant for
understanding kinetic processes (among hard on-shell particles)
occuring right after a heavy-ion collision.

\vspace*{1cm}
\begin{acknowledgments}
A.D.\ gratefully acknowledges support by the DOE Office of Nuclear
Physics through Grant No.\ DE-FG02-09ER41620; and from The City
University of New York through the PSC-CUNY Research Award Program,
grants 65041-00~43 and 66514-00~44.

\end{acknowledgments}


\end{document}